\documentclass[showpacs,pre,twocolumn]{revtex4}
\makeatletter
\newif\ifcolor
\colortrue
\def\place@bibnumber@inl#1{#1.}%

\makeatother

\usepackage{here}
\usepackage{epsf}
\usepackage{amsmath}
\usepackage{times}
\def\Eq#1{Eq.~(\ref{#1})}
\def\Eqs#1{Eqs.~(\ref{#1})}
\def\half{\mbox{$1\over 2$}}
\def\({\left (}
\def\){\right )}
\def\[{\left [}
\def\]{\right ]}

\def\ds{\displaystyle}

\def \pra#1,#2,{{\partial #1\over \partial #2}}
\def \parb#1,#2,#3,{{\partial #1\over \partial #2}\left.\right\vert_{#3}}
\def \parc#1,#2,#3,{{\partial^2 #1\over\partial #2\partial #3}}
\def \pard#1,#2,{{\ds\partial^2 #1\over\ds \partial #2^2}}
\def\<{\left \langle}
\def\>{\right\rangle}
\begin{document}
\title[Geometry of flowers]{Geometry and elasticity of strips and flowers}
\author{M. Marder} 
\affiliation{Center for Nonlinear Dynamics and Department of Physics
The University of Texas at Austin, Austin TX 78712, USA}
\email{marder@chaos.ph.utexas.edu}
\author{N. Papanicolaou} 
\affiliation{Department of Physics,  University of Crete, and
  Research Center of Crete, Heraklion, Greece} 
\email{papanico@physics.uoc.gr}
%\date{\today}
\begin{abstract}
  We solve several problems that involve imposing metrics on surfaces.
 The problem of a strip with a linear metric gradient
   is formulated in terms of a Lagrangean similar to those used
   for spin systems. We are able to show that the low energy state
  of long strips is a twisted helical state like a telephone cord. We
  then extend the techniques used in this solution to two--dimensional
  sheets with more general metrics. We find evolution equations and
  show that when they are not singular, a surface is determined by
  knowledge of its metric, and the shape of the surface along one
  line.  Finally, we provide numerical evidence that once these
  evolution equations become singular, either the surface is not
  differentiable, or else the metric deviates from the target metric
  as a result of minimization of a suitable energy functional .
\end{abstract}
\maketitle

\section{Introduction}

In a series of experiments performed in 2002, Sharon {\it et.
  al.}\cite{Sharon.02} showed that thin sheets deformed in a smooth
fashion spontaneously fold into convoluted shapes with much less
symmetry than the original deformation. These experiments raised many
questions about the relationship between local changes in distance
within a sheet, and the global shape the sheet adopts. The first
theoretical papers that set about explaining the
experiments\cite{Marder.strip.02,Audoly.02,Marder.FOP.03,Marder.EPL.03,Audoly.03b}
established certain points, but left many others unsettled.
What was most clearly established was that one should study the
problem by choosing a metric function suggested by the physical
process that deformed the sheet, and trying to determine what surfaces
were compatible with this metric.  Much of the effort in these papers
was devoted to exploring the problem of a long thin strip of material
with a linear metric gradient in one direction. This strip problem was
not solved in great generality, and its relation to the more general
two--dimensional problems was not definite. Furthermore, the basic idea
of the calculations was not clear. On the one hand, the papers spoke
of imposing a metric on a sheet, while on the other there were many
indications that the actual metric of the final shape adopted by the
sheet was something else.

This paper has three main sections which address the questions
we have just described.

 First, we return to the problem of
the long strip with a metric gradient and provide a  more complete
solution than was afforded previously. In particular, we find the
lowest--energy shape of this strip when it becomes very long. In this
limit, the strip coils into a helix like a telephone cord. The
edge of a flower or a leaf cannot behave in this way. Therefore,
solutions of the strip problem have only limited relevance to the more
general two--dimensional questions raised by the experiments. However,
the formal techniques used to solve the strip prove to be helpful, and
provide the basis for our approach to the two--dimensional problem.

Second, we return to the problem of two--dimensional sheets, and
attempt to determine the extent to which specifying a metric
determines the surface. We can discard the possibility that a metric
alone can dictate the shape of a surface. For example, a piece of
paper can be bent into infinitely many smooth shapes, all of which
share the same Euclidean metric.  This property is not simply due to the fact
that the piece of paper is initially flat. Tape the paper into the
shape of a cylinder, and still it can be deformed in infinitely many
ways without change of metric. A classic theorem of differential
geometry states that surfaces are uniquely determined, up to rotations
and translations, by the metric, and by the second fundamental form,
provided that they satisfy the Gauss--Codazzi equations\cite{Eisenhart.59}.
 However, the second fundamental form provides
more information than is needed. We will show that if one knows the
metric of a surface, and the shape of the surface along one line, then
the rest of the surface is determined by an evolution equation.  This
statement must be qualified. The procedure that determines the surface
typically contains singularities, and the attempt to find the surface with
geometry alone comes to a halt.

Third, we discuss numerical procedures to find surfaces corresponding
to metrics for which the evolution equations of the previous section
become singular. We describe in some detail an energy functional whose
minima should provide surfaces that approach a desired target metric
as closely as possible. In cases where evolution equations dictate
surfaces based upon geometry alone, the energy functional recovers
them. When the evolution equations fail to find a surface, then either
the energy functional finds a non-differentiable surface, or else it
finds a smooth flower--like surface whose metric is different from the
one that was supposedly imposed.

\section{\label{sec:strip}Ground state for twisted strips}
%\subsection{Motivation}

Reference \cite{Sharon.AmSci.04} shows a number of thin strips cut
from the edge of a leaf. Each of them curls up into a circle, with the
curvature of each strip depending upon the gradient of the metric at
that point in the leaf. Such observations suggested that the rippling
pattern at the edge of a leaf could be understood by focusing upon a
thin strip of material with a metric gradient. Such a strip, freely
allowed to seek out its lowest energy state simply curls up into a
circle. Therefore, the additional constraint was added that the two
ends of the strip had to be some distance $\lambda$
apart\cite{Marder.strip.02}. This problem yielded nontrivial
solutions\cite{Marder.strip.02,Audoly.02,Marder.EPL.03,Marder.FOP.03}.
Certain special cases could be found analytically. Somewhat more
general solutions could be found numerically. However, the numerical
procedures were not very stable, and left open a number of
questions. For example, there was some speculation that as the length
$L$ of the curved strip became infinite with $L/\lambda$ kept finite,
the lowest--energy state might be fractal. This speculation was
incorrect, as we now show through a correct solution of the problem.
\subsection{Problem setting}

\begin{figure}
\begin{center}
\epsfxsize\columnwidth\epsffile{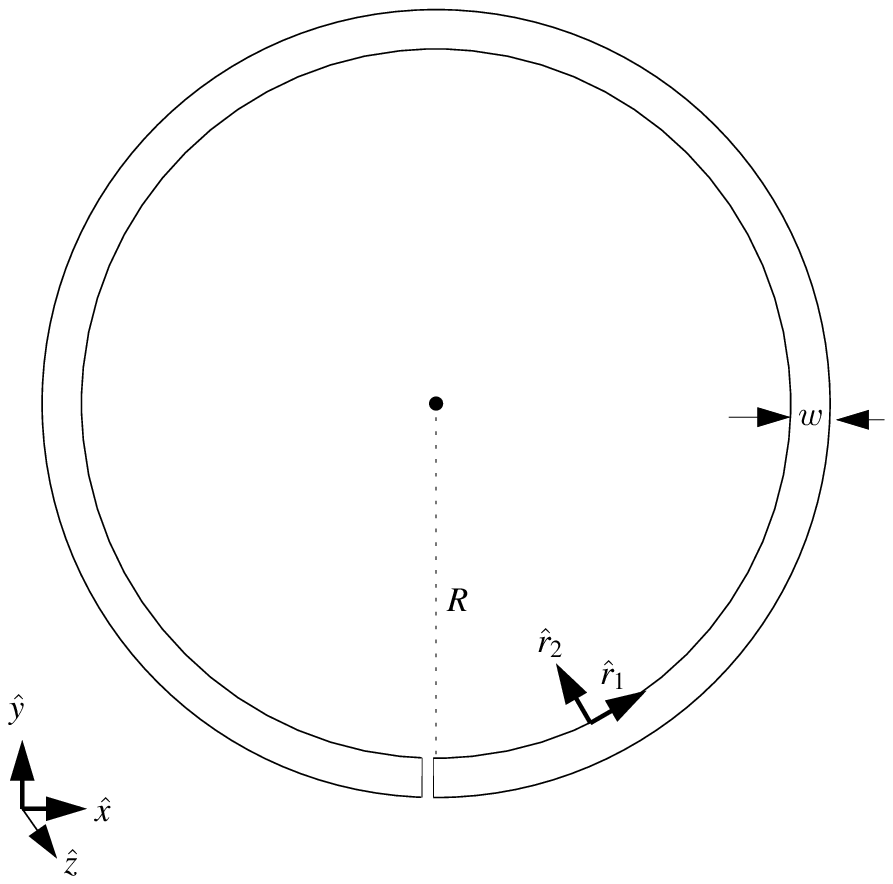}
\caption{\small
  The buckled strips studied in this section can be examined by
  cutting out this circular strip, of radius $R$ and width $w$.
  Roughly speaking, the main question addressed here is to predict the
  shape of the strip when the two ends are pulled apart to some
  specified distance and relative orientation.  For the general
  solutions studied in this section, the total length $L$ of the strip
  need not equal $2\pi R$. }
\label{fig:circle}
\end{center}
\end{figure}

We briefly review the specification of the strip problem.  Consider a
circular strip of paper, as shown in Figure \ref{fig:circle}, of
radius $R$, length $L=2\pi R$, width $w$, and thickness $t$. The
length $L$ can be arbitrary, and may be greater or less than $2\pi R$.
We will be interested in strips where the following limits apply:
\begin{equation}
\left ({w\over R}\right )^2\ll {t\over R}\ll {w\over R}\ll {L\over R}.
\label{eq:inequalities}
\end{equation}
When these conditions hold, the energy of the strip takes a very simple
form as a functional of a line passing through the center of the
strip\cite{Marder.strip.02}. 

In particular, let $s$ be the arc length along the center of the strip.  The
low--energy conformations of the strip are captured by two orthonormal vectors
$\hat r_1(s)$ and $\hat r_2(s)$, where $\hat r_1$ points along the arc $s$,
and $\hat r_2$ lies in the plane of the strip and is orthogonal to
$\hat r_1$. Define also 
\begin{math}
\hat r_3=\hat r_1\times\hat r_2,
%\label{eq:r3}
\end{math}
so as to obtain a set of unit vectors that describes in a natural way
the local orientation of the strip. The location $\vec r(s)$ of each point along the
center--line  of the strip can be obtained from 
\begin{equation}
\vec r(s)=\int_0^s ds'\,\hat r_1(s').
\label{eq:l}
\end{equation}
At each point $s'$ along the way, the precise orientation of the strip is
specified by the unit vector $\hat r_2$, or equivalently by the unit
vector $\hat r_3,$ which is normal to the strip.
Therefore, the vectors $\hat r_1(s)$ and $\hat r_2(s)$ contain all the
information necessary to deduce the shape of the strip when it is
embedded in three--dimensional space. The inequalities
(\ref{eq:inequalities}) ensure that the strip remains essentially flat
and featureless in the direction pointing along $\hat r_2$.

In terms of the vectors $\hat r_1(s)$ and $\hat r_2(s)$ the energy of
the strip is given by\cite{Marder.strip.02,Audoly.02}
\begin{equation}
U=\int_0^L ds\,{\cal E}(s),
\end{equation}
where
\begin{equation}
{\cal E}={C_1\over2}\left(|\hat
  r_1'|^2-1/R^2\right)+{C_2\over 2}\left(|\hat r_2'|^2-1/R^2\right ),
\label{eq:energy1}
\end{equation}
and primes indicate derivatives with respect to arc length $s$.

This energy is subject to two constraints. The first is a {\it local
  constraint} that results from the
fact that the strip has a metric gradient in the direction of $\hat r_2$ that
causes it to want to curl up with radius of curvature $R$. This
constraint is captured by the condition
\begin{equation}
\hat r_1'\cdot\hat r_2 =\ -\hat r_2'\cdot\hat r_1=1/R.
\label{eq:local_constraint}
\end{equation}
The second constraint is that the ends of the strip be pulled apart by
some distance; this constraint prevents the strip from simply curling
up into a circle. We impose this constraint through
\begin{equation}
(\vec r(L)-\vec r(0))\cdot\hat z=\lambda.
\label{eq:global_constraint}
\end{equation}
That is, the $z$ component of the difference between starting and
ending points of the strip is constrained to have value $\lambda$. By
scanning through values of $\lambda$, one equivalently scans through
all allowed distances between the starting and ending point of the
strip. The constraint in the form of \Eq{eq:global_constraint} is much
easier to work with formally than if it were literally expressed in
terms of end--to--end distance.  For definiteness, we choose a fixed
(laboratory) frame as shown in Figure \ref{fig:circle}. The $\hat z$
axis is taken to be perpendicular to the plane defined by the initial
(undeformed) circular strip, while the $\hat x$ and $\hat y$ axes
point along the initial directions of $\hat r_1$ and $\hat r_2$ at one
of the endpoints of the strip.

\def\cpsi{\cos\psi}
\def\spsi{\sin\psi}
\def\cphi{\cos\phi}
\def\sphi{\sin\phi}
\def\cth{\cos\theta}
\def\sth{\sin\theta}

The trihedral $\hat r_1, \hat r_2, \hat r_3$ uses nine coordinates to
represent the orientation of the strip, when in fact only three are
needed.
\begin{widetext}
Further analysis is greatly simplified by rewriting the unit
vectors in terms of Euler angles (Ref.\cite{Goldstein.69} Eq.4.46), through

\begin{equation}
\begin{array}{rllll}
\hat r_1&=&[\phantom{-}\cpsi\cphi-\cth\sphi\spsi,& \phantom{-}\cpsi\sphi+\cth\cphi\spsi, &\sth\spsi];\\
\hat r_2&=&[
-\spsi\cphi-\cth\sphi\cpsi,
\ \ &  -\spsi\sphi+\cth\cphi\cpsi,\
\ &\sth\cpsi];\\
\hat r_3&=&[\phantom{-}\sth\sphi & -\sth\cphi,& \cth].
\end{array}
\label{eq:eangles}
\end{equation}
\end{widetext}

With this representation, we can rewrite the local constraint
\Eq{eq:local_constraint} as
\begin{equation}
 \cos \theta\  \phi'+ \psi'=  1/R;
\label{eq:local_constraint1}
\end{equation}
again, primes refer to derivatives with respect to the arc length
$s$. 

Writing out  \Eq{eq:energy1} in terms of the Euler angles  one has 
\begin{equation}
\begin{array}{lll}
{\cal E}&=&\half\left [ (C_2-C_1)\cos^2\psi+C_1\right]\theta'^2\\[5pt]
&&+( C_2-C_1)\cos\psi \sin\psi
\sin\theta\ \phi'\theta'\\[5pt]
&&+\half \left [(C_1-C_2)\cos^2\psi+C_2\right]\sin^2\theta\ \phi'^2.
%&&  -h\sin(\psi)\sin(\theta).
\end{array}
\label{eq:Lagrangean1}
\end{equation}
In order to impose the global constraint \Eq{eq:global_constraint}, we
will employ a Lagrange multiplier, and write down the Lagrangean
\begin{equation}
\begin{array}{lll}
{\cal L}&=&\half\left  [ (C_2-C_1)\cos^2\psi+C_1\right]\theta'^2\\[5pt]
&&+( C_2-C_1)\cos\psi \sin\psi
\sin\theta\ \phi'\theta'\\[5pt]
&&+\half \left [(C_1-C_2)\cos^2\psi+C_2\right]\sin^2\theta\ \phi'^2\\[5pt]
&&  -h\sin\psi\sin\theta.
\end{array}
\label{eq:Lagrangean2}
\end{equation}
Note that $\phi'$ appears in \Eq{eq:Lagrangean2}, but not $\phi$
itself. This fact will allow for further simplification.

The constants $C_1$ and $C_2$ depend upon elastic properties of the
strip. For a particular case studied in Ref.\cite{Marder.FOP.03},
they are related through $2C_2=3 C_1$. Materials with different
elastic properties would give different constants, so we take $C_1$
and $C_2$ as free variables. The structure 
of the equations is particularly simple when $C_1=C_2,$ so we will
begin with that case, and return later to the more general situation.

Adopting $C_1=C_2=1$, the Lagrangean of \Eq{eq:Lagrangean2} simplifies to
\begin{equation}
{\cal L}={1\over 2} \theta'^2+{1\over 2} \phi'^2
\sin^2\theta-h\sin\psi\sin\theta.
\label{eq:Lagrangean_simple1}
\end{equation}
All appearances of $\phi$ can now be eliminated by employing the local
constraint \Eq{eq:local_constraint1} to give
\begin{equation}
{\cal L}={1\over 2} \theta'^2+{1\over 2} (\psi'-1/R)^2
\tan^2\theta-h\sin\psi\sin\theta.
\label{eq:Lagrangean_simple2}
\end{equation}
Similar Lagrangeans appear in the study of spin
systems\cite{Chovan.02}. 
The equations of motion following from \Eq{eq:Lagrangean_simple2} are
\begin{subequations}
\begin{eqnarray}
 \Big [ (\psi'-1/R)\tan^2\theta
\Big]'=-h\cos\psi\sin\theta
\label{eq:eq_motion1;a}
\\
\theta''=(\psi'-1/R)^2{\sin\theta\over\cos^3\theta}-h\sin\psi\cos\theta
\label{eq:eq_motion1;b}
\end{eqnarray}
\label{eq:eq_motion1}
\end{subequations}

\subsection{Particular solution}
Ref. \cite{Marder.strip.02} provided a family of exact solutions, and
we search for these again with this new 
formalism. We find that they emerge if one sets $h=0$  in
\Eqs{eq:eq_motion1}. In this case,
one  immediately integrates
\Eq{eq:eq_motion1;a} to obtain
\begin{equation}
(\psi'-1/R)\tan^2\theta=\beta,
\end{equation}
where $\beta$ is an integration constant.
Inserting this relation into \Eq{eq:eq_motion1;b}
and integrating gives
\begin{eqnarray}
\nonumber \theta'&=&\sqrt{\alpha^2-\beta^2/\sin^2\theta}\\
\Rightarrow\cos\theta&=&-\sqrt{1-(\beta/\alpha)^2}\sin(\alpha s),
\label{eq:cos_th_ans}
\end{eqnarray}
where $\alpha$ is an additional integration constant. Without loss of
generality one can choose $\alpha>0$. Note from \Eq{eq:cos_th_ans}
that $\alpha>\beta$ if the solutions are to remain real.

For $\psi$ one obtains
\begin{equation}
\psi=\tan^{-1}\Big((\beta/\alpha)\tan(\alpha s)\Big)-(\beta-1/R)
s+\psi_0.
\label{eq:psi_ans}
\end{equation}
Similarly, for $\phi$ one obtains
\begin{equation}
\phi=-\tan^{-1}\Big(\sqrt{(\alpha/\beta)^2-1}\cos(\alpha
s)\Big)+\phi_0;
\label{eq:phi_ans}
\end{equation}
the integration constants $\psi_0$ and $\phi_0$ give the value of
$\psi$ and $\phi$ when $s=0$.

The global constraint \Eq{eq:global_constraint} requires that 
\begin{equation}
\int_0^L ds\,\sin\theta\sin\psi=\lambda.
\label{eq:global_constraint2}
\end{equation}
As $\lambda$ becomes large, it is only possible to satisfy
\Eq{eq:global_constraint2} if $\sin\theta$ and
$\sin\psi$ have the same period. Comparing \Eqs{eq:psi_ans} and
(\ref{eq:cos_th_ans}), one sees that this condition can be satisfied by
taking 
\begin{equation}
\beta=\alpha+1/R.
\end{equation}
With this choice, one has
\begin{equation}
\sin\theta\sin\psi=\hat r_1\cdot\hat z=(\alpha R+\sin^2(\alpha s))/(\alpha R),
\end{equation}
which with a bit of manipulation upon setting $R=1$ reproduces Eq.
(41a) in Ref. \cite{Marder.strip.02}; note that the $x$ axis in that
reference corresponds to the $z$ axis here. The remainder of the
special solution obtained previously can be recovered as well. Since
this special solution depends upon setting $h=0$, or equivalently to
fixing a relationship between $\lambda$ and $L$ not required by the
original problem, we move to a numerical approach capable of solving
the problem in greater generality.
\subsection{Numerical solutions}

\begin{figure}[!b]
\epsfxsize=\columnwidth\epsffile{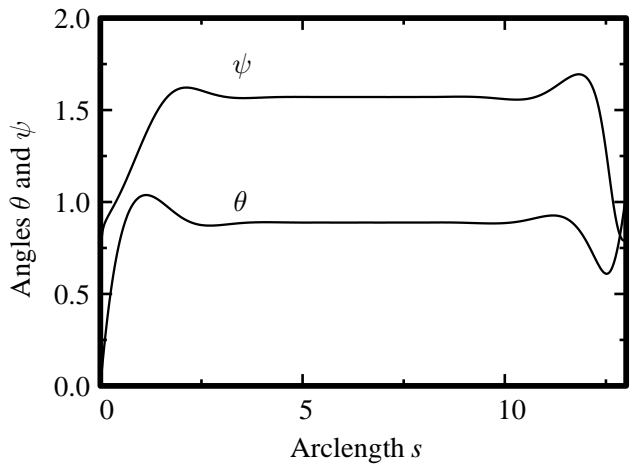}
\caption{Plot of Euler angles $\theta$ and $\phi$ for system with
  total length $L=13$, constrained to have height in $\hat z$
  direction $\lambda=.75L$. Note that apart from some
  variation near the endpoints needed to obey the boundary conditions,
  both angles go to constant values throughout most of the length of
  the sample.}
\label{fig:euler_angles_13_.75}
\end{figure}

We can obtain reliable numerical ground states of
\Eq{eq:Lagrangean_simple1}  by
changing the form of the Lagrange multiplier to enforce the global
constraint in a fashion that involves a positive definite functional: 
\begin{eqnarray}
\nonumber F&=&\int_0^L ds\,\left [ {1\over 2} \theta'^2+{1\over 2}
  (\psi'-1/R)^2 
\tan^2\theta \right ] \\&+&H\left(\lambda-\int_0^Lds\,\sin\psi\sin\theta\right)^2,
\label{eq:Lagrangean_simple3}
\end{eqnarray}
where $H$ is chosen on the order of $100$. To find solutions, we
simply minimize $F$.

\begin{figure}[!bhtp]
  \centering
  \ifcolor
  (a)\epsfysize=3.5in\epsffile{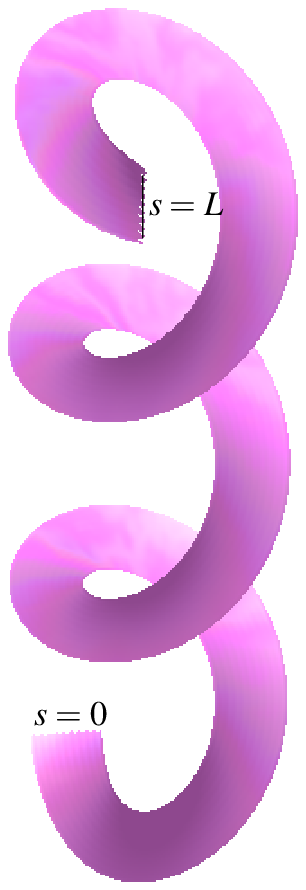} \ \ \ \ \ \ \ 
  (b)\epsfysize=3.5in\epsffile{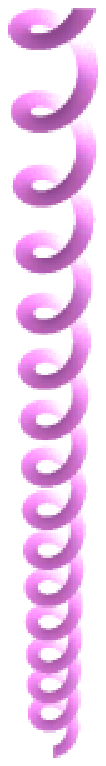}
  \else
  (a)\epsfysize=3.5in\epsffile{euler_visualize_labeled_bw.eps} \ \ \ \ \ \ \ 
  (b)\epsfysize=3.5in\epsffile{strip_visualize_bw.eps}
  \fi
  \caption{(a) Three--dimensional visualization of solution shown in
  Figure  \ref{fig:euler_angles_13_.75}\ifcolor (in color)\fi. (b)
    Visualization of low--energy 
    solution for a long strip with $\lambda/L=0.625.$ The strip winds in
    helical fashion around the $\hat z$ axis \ifcolor (in color)\fi.}
  \label{fig:strip_visualize}
\end{figure}

In the minimization procedure we fix boundary conditions
on $\theta$ and $\psi$ at $s=0$ and $s=L$. We may also choose the
initial ($s=0$) value of $\phi$, but its final ($s=L$) value is left
free to adjust according to Eq. (\ref{eq:local_constraint1}).   Now
we keep one of the endpoints of the strip fixed and
orient corresponding initial directions of
$\hat r_1$ and $\hat r_2$ along the $\hat x$ and $\hat y$ axes
of the laboratory frame. Noting that when $\theta=0$, the orientation of
the trihedral depends only upon $\phi+\psi$, the initial values of the Euler angles
are set at
\begin{equation}
          \theta(0)=0,\ \psi(0)+\phi(0)=0.
\label{eq:initial_angles}
\end{equation}
At the other endpoint, the values of $\theta$ and $\psi$ are chosen arbitrarily;
    e.g.,
\begin{equation}
  \theta(L)=\pi/3,\  \psi(L)=\pi/4,
\label{eq:final_angles}
\end{equation}
while the end--to--end distance is controlled by the parameter $\lambda$.  A 
numerical solution for $L=13$ and 
$\lambda=0.75L$ yields values for $\theta$ and $\psi$ shown in Figure
\ref{fig:euler_angles_13_.75}. The angle $\phi=\phi(s)$ is obtained by
an elementary integration of \Eq{eq:local_constraint1}, and the
complete orthonormal trihedral from \Eq{eq:eangles}. It is then
straightforward to calculate $\vec r=\vec r(s)$ from \Eq{eq:l}
which determines the relative position of the strip in
three-dimensional space, as illustrated in Figure
\ref{fig:strip_visualize} (a). 

One property of the solution that might seem unsatisfactory is that
the strip wraps a number of times around the $\hat z$ axis. If one
were to grab two ends of a strip and pull them apart, this would not
be allowed. If one wants to find energy--minimizing solutions without
any net twist around the vertical axis, the following strategy is
effective: 
Take a solution obeying the boundary conditions 
\begin{equation}
\theta(L)=\pi/2,\quad \psi(L)=\pi/2,
\label{eq:bc1}
\end{equation}
produce a mirror image with $z\rightarrow -z$, $x\rightarrow x$,
$y\rightarrow y$. The resulting function also minimizes the functional
$F$. It can be
glued on to the solution found so far, joining smoothly to it at
$s=L$ because of \Eq{eq:bc1}, and will continue on to terminate at $(0,0,2\lambda)$ when
$s=2L$. The second half of the solution reverses the twist produced by
the first half. In all cases we have checked, solutions of this type are
the lowest--energy solutions without net twist.
  
A notable feature of Figure \ref{fig:euler_angles_13_.75} and others
like it is that the angles 
$\theta$ and $\psi$ quickly approach constant values away from the
endpoints, which do not depend on the specific boundary conditions
(\ref{eq:initial_angles}) and (\ref{eq:final_angles}). Thus we
conclude that for {\it long strips}, the energy
minimizing solutions have the following properties:
 $\theta$ and
$\psi$ are constants such that
\begin{equation}
\sin\psi\sin\theta={\lambda/L},
\label{eq:ST10}
\end{equation}
in order to satisfy the global constraint.
According to  \Eq{eq:eq_motion1;a}, $\cos\psi\sin\theta=0$, which is
compatible with \Eq{eq:ST10} only if 
\begin{equation}
\psi=\pi/2, \quad \sin\theta={\lambda/L}.
\label{eq:uniform_solution}
\end{equation}
For solutions obeying \Eq{eq:global_constraint2}, the corresponding energy density
\begin{equation}
{\cal E}={1\over 2R^2}\tan^2\theta={1\over 2 R^2}{\lambda^2\over
  L^2-\lambda^2}
\end{equation}
is also constant, independent of $s$. Nevertheless, the solution
oscillates because of \Eq{eq:local_constraint1}, which implies
that
\begin{equation}
\phi={s\over R\cos\theta}={s\over R\sqrt{1-(\lambda/L)^2}}.
\end{equation}
The period of oscillation is given by 
\begin{equation}
s_{\rm osc}=2\pi R\cos\theta=2\pi R\sqrt{1-(\lambda/L)^2}.
\end{equation}
and the distance the solution travels when $s$ traverses this arc-length is
\begin{equation}
z_{\rm osc}=2\pi R\sin\theta\cos\theta=2\pi
R\sqrt{1-(\lambda/L)^2}{\lambda\over L}.
\end{equation}
This last expression gives the period of oscillation one would
measure upon taking a strip with $L\gg R$ and pinning its ends at
distance $\lambda$ in the laboratory. A three--dimensional
visualization of such a strip appears in Figure \ref{fig:strip_visualize}(b).

We conclude that long strips with metric gradients minimize their
energy by curling up like telephone cords. They deviate from this
quasi--uniform solution only when they must maneuver near the ends to obey boundary
conditions, or form a kink in the middle to produce a solution with no
net twist.

Returning to the general case where $C_1\neq C_2$ and writing out the
Euler--Lagrange equations, we find that the uniform configuration
(\ref{eq:uniform_solution}) is still a solution. The energy density is
given by ${\cal E}=C_2\tan^2\theta/2R^2$ and all other features of the
  global minimum such as the period of oscillation remain unchanged.

\def\F{F}
\def\G{G}
\def\y{y}
\def\x{x}

\section{Geometrical considerations\label{sec:Geometry}}

The strip problem  was originally
motivated by experiments on two--dimensional sheets of deformed
material. The expectation was that by focusing upon a long thin strip,
one could understand basic geometrical features of the full
two--dimensional problem. Now that we have obtained a reasonably detailed account
of the ground state of the strip, it is clear that its significance
for the original problem is limited. Long thin strips wrap up in a
helix, like a telephone cord. This geometry is impossible for the
edges of a flower or a leaf. Therefore, in Section \ref{sec:Numerics} we return to the
two--dimensional problem and search for some alternate approaches.
 As preparation, this section summarizes some
basic facts concerning the differential geometry of surfaces.

Let  $\vec r=(x,y,z)$
be a point in three--dimensional Euclidean space, where $x$, $y$, and
$z$ are the usual Cartesian coordinates with respect to a fixed
laboratory frame. A surface may then be defined in parametric form by 
\begin{equation}
\vec r=\vec r(u,v)=\Big (x(u,v),y(u,v),z(u,v)\Big)
\label{eq:r2}
\end{equation}
where $u$ and $v$ are parameters whose range is not specified for the
moment. Throughout this section, derivatives will be abbreviated by
$\partial_1={\partial/ \partial u}$ and $\partial_2={\partial/ \partial v}$.
The elements of the metric tensor are then defined from
\begin{equation}
  \label{eq:MetricDefine}
  g_{\alpha\beta}=(\partial_\alpha\vec r)\cdot(\partial_\beta\vec r).
\end{equation}

A second fundamental form (tensor) is defined as follows: Let $\hat
r_3=\hat r_3(u,v)$ be the unit vector that is perpendicular to the
surface at point $(u,v)$ and varies continuously with $u$ and $v$. The
elements of the second fundamental form are then given by
\begin{equation}
  \label{eq:SecondForm}
  d_{\alpha\beta}=\hat r_3\cdot\partial_\alpha\partial_\beta\vec r.
\end{equation}
The two symmetric tensors $g_{\alpha\beta}$ and $d_{\alpha\beta}$ play
a significant role in the theory of surfaces, as is apparent in
standard texts\cite{Eisenhart.59,Pogorelov.56}.

\subsection{Trumpets\label{sec:trumpet}}

In order to gain familiarity with the types of surfaces studied in
Section \ref{sec:Numerics}, we first consider an elementary example
defined by

\begin{equation}
\label{eq:Trumpet0}
\vec r=\rho(v)(\cos u \hat e_1+\sin u\hat e_2)+\zeta(v)\hat e_3, 
\end{equation}
where $\hat e_1$, $\hat e_2$, and $\hat e_3$ are constant unit vectors
along the three axes of the laboratory frame, while $\rho(v)$ and
$\zeta(v)$ are functions of $v$ alone, and are further restricted by
the condition
\begin{equation}
  \label{eq:restriction}
  \rho'^2+\zeta'^2=1,
\end{equation}
where the primes indicate derivatives with respect to $v$. The surface
is thus specified by the single function $\rho=\rho(v)$. In
particular, the metric tensor is then given by
\begin{equation}
  \label{eq:TrumpetMetric}
  g_{11}=\rho^2,\ g_{22}=1,\  g_{12}=0.
\end{equation}
In order to calculate the second fundamental form, we first construct
the orthonormal trihedral
\begin{eqnarray}
\nonumber\hat r_1&=&{1\over \rho}\partial_1\vec r=-\sin u\hat e_1+\cos u\hat
e_2 \\
\hat r_2&=&\partial_2\vec r=\rho'(\cos u\hat e_1+\sin u\hat
e_2)+\sqrt{1-\rho'^2}\hat e_3 \label{eq:TrumpetTrihedral}\\[3pt]
\nonumber\hat r_3&=&=\sqrt{1-\rho'^2}(\cos u \hat
e_1+\sin u \hat e_2)-\rho'\hat e_3,
\end{eqnarray}
where $\hat r_1$ and $\hat r_2$ are tangent to the surface, while
$\hat r_3=\hat r_1\times\hat r_2$ is perpendicular. The second fundamental form is then
calculated from \Eq{eq:SecondForm} to yield
\begin{eqnarray}
\nonumber d_{11}&=&-\rho\sqrt{1-\rho'^2},\ d_{22}={\rho''\over\sqrt{1-\rho'^2}}\\
d_{12}&=&d_{21}=0
\label{eq:TrumpetSecondForm}
\end{eqnarray}

A special case of this class of surfaces is the ordinary cylinder with
unit radius, obtained with the choice $\rho=1$ and $\zeta=v$. The
metric is then Euclidean ($g_{11}=1=g_{22}, g_{12}=g_{21}=0$) and the
elements of the second fundamental form are $d_{11}=-1, d_{22}=0,
d_{12}=d_{21}=0$. 

\begin{figure}[htbp]
  \centering
  \ifcolor
    \epsfxsize=3.3in\epsffile{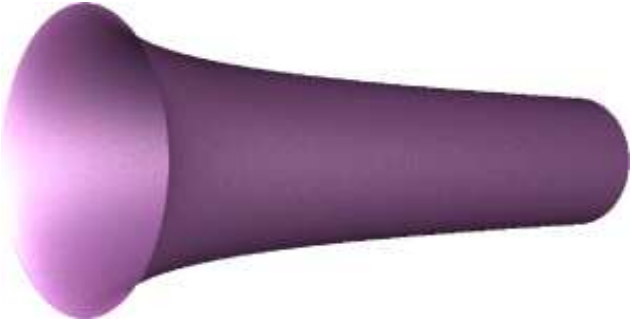}
    \else
    \epsfxsize=3.3in\epsffile{Roussos04-07-26__3visualize.eps}
\fi
\caption
 {Rendering of the surface produced by the metric in \Eq{eq:Trumpet}
 for $u\in[0,2\pi]$,and $v\in(-\infty,0]$ \ifcolor (in color)\fi.}
  \label{fig:Trumpet}
\end{figure}

A more interesting example is obtained by the choice
\begin{equation}
  \label{eq:Trumpet}
  \sqrt{g_{11}}=\rho=1+{1\over 1-v},
\end{equation}
which leads to an axially symmetric surface with variable radius
$\rho=\rho(v)$ and height $z=\zeta(v)$ calculated from
\Eq{eq:restriction}. This surface reduces to the ordinary cylinder in
the limit $v\rightarrow -\infty$, but its radius increases with
increasing $v$ (or $z$). It would appear that the radius of this
surface would eventually grow to infinity as $v\rightarrow
1$. Actually, this limit cannot be obtained because of a singularity
that develops earlier when $\rho'^2=1$ or $v=0$. The actual surface is
illustrated in Figure \ref{fig:Trumpet}; we will refer to
it as the {\it trumpet}. It starts as a cylinder with radius $\rho=1$
and terminates at a cusp with radius $\rho=2.$ Near the cusp the
elements of the second fundamental form calculated from
\Eq{eq:TrumpetSecondForm} approach the characteristic limits
$d_{11}\rightarrow 0, d_{22}\rightarrow\infty$, while $d_{12}$
vanishes everywhere. 

This trumpet may be considered the most primitive model of a
flower. However, a true flower is a trumpet that is allowed to grow beyond the
cusp, and displays numerous ripples, because biology and elasticity
dictate slight modifications of the metric and second fundamental form
needed to evade the singularity. Geometry alone cannot resolve what
happens as one tries to push a surface beyond such a cusp, but may
help to classify the possibilities.

Hence, in the remainder of this section, we formulate an inverse
problem of sorts that will enable us to interpret the explicit results
of Section \ref{sec:Numerics}. 

\subsection{Compatibility conditions}
The elementary example we have just
considered is significantly generalized by considering surfaces for
which the coordinate curves form an {\it orthogonal net}, which means
that they are characterized by a metric of the form
\begin{equation}
  \label{eq:RoussosMetric}
  g_{11}=g_{11}(u,v),\ g_{22}=g_{22}(u,v),  \ g_{12}=0.
\end{equation}
The corresponding orthonormal trihedral is then defined from
\begin{equation}
  \label{eq:RoussosTrihedral}
  \hat r_1={\partial_1\vec r\over\sqrt{g_{11}}},\ \hat
  r_2={\partial_2\vec r\over \sqrt{g_{22}}},\ \hat r_3=\hat r_1\times
  \hat r_2.
\end{equation}

We now consider the elementary integrability condition
\begin{eqnarray}
\nonumber  
  \partial_1\partial_2\vec r&=&\partial_2\partial_1\vec r\\
  \label{eq:RoussosCompatibility}
\Rightarrow \partial_1(\sqrt{g_{22}}\hat
r_2)&=&\partial_2(\sqrt{g_{11}}\hat r_1)\\
\nonumber
\Rightarrow(\partial_1\sqrt{g_{22}})\hat
r_2+\sqrt{g_{22}}\partial_1\hat r_2&=&(\partial_2\sqrt{g_{11}})\hat r_1
+\sqrt{g_{11}}\partial_2\hat r_1\ \ 
\end{eqnarray}
Contracting both sides of \Eq{eq:RoussosCompatibility} with the unit
vectors $\hat r_1$, $\hat r_2$ and $\hat r_3$ in turn we obtain
\begin{subequations}
\label{eq:RoussosEqs}
\begin{eqnarray}
\label{eq:RoussosEqs;a}
(\hat r_1\cdot\partial_1\hat
r_2)&=&{\partial_2\sqrt{g_{11}}\over\sqrt{g_{22}}}\\
\label{eq:RoussosEqs;b}
(\hat r_2\cdot\partial_2\hat
r_1)&=&{\partial_1\sqrt{g_{22}}\over\sqrt{g_{11}}}\\
\label{eq:RoussosEqs;c}
\sqrt{g_{11}}(\hat r_3\cdot\partial_2\hat r_1)&=&\sqrt{g_{22}}(\hat
r_3\cdot\partial_1\hat r_2).
\end{eqnarray}
\end{subequations}
We observe that for $\sqrt{g_{11}}=1-v/R$, $g_{22}=1$,
\Eq{eq:RoussosEqs;a} reproduces \Eq{eq:local_constraint}. As in
Section \ref{sec:strip}, we parameterize the trihedral with the 
Euler angles of \Eq{eq:eangles} to obtain the
three fundamental equations
\begin{subequations}
\label{eq:RoussosEqs2}
\begin{eqnarray}
\cos\theta\,\partial_1\phi+\partial_1\psi&=&-{\partial_2\sqrt{g_{11}}\over
  \sqrt{g_{22}}} \label{eq:RoussosEqs2;a}\\
\cos\theta\,\partial_2\phi+\partial_2\psi  &=&{\partial_1\sqrt{g_{22}}\over
  \sqrt{g_{11}}}
\label{eq:RoussosEqs2;b}
\end{eqnarray}
\begin{eqnarray}
\nonumber
\sqrt{g_{11}}(&-&\sin\theta\cos\psi\,\partial_2\phi+\sin\psi\,\partial_2\theta)\\
&=&\sqrt{g_{22}}(\sin\theta\sin\psi\,\partial_1\phi+\cos\psi\partial_1\theta),
\label{eq:RoussosEqs2;c}
\end{eqnarray}
\end{subequations}
which will provide the basis for subsequent development. It should be
noted that the compatibility  conditions \Eqs{eq:RoussosEqs2} are
formulated entirely in terms of the metric tensor $g_{\alpha\beta}$
and that the elements of the second fundamental form do not appear
explicitly. In fact, once a solution of \Eqs{eq:RoussosEqs2} is
available, $d_{\alpha\beta}$ can be computed from \Eq{eq:SecondForm} as
\begin{subequations}
\begin{eqnarray}
\nonumber d_{11}&=&\sqrt{g_{11}}(\hat r_3\cdot\partial_1\hat
r_1)\\
&=&\sqrt{g_{11}}\Big(-\sin\theta\cos\psi\,\partial_1\phi
+\sin\psi\,\partial_1\theta\Big )\\
\nonumber d_{22}&=&\sqrt{g_{22}}(\hat r_3\cdot\partial_2\hat
r_2)\\
&=&\sqrt{g_{22}}\Big(\sin\theta\sin\psi\,\partial_2\phi
+\cos\psi\,\partial_2\theta\Big )\\
\nonumber 
d_{12}&=&\sqrt{g_{22}}(\hat r_3\cdot\partial_1\hat
r_2)\\
&=&\sqrt{g_{22}}\Big(\sin\theta\sin\psi\,\partial_1\phi
+\cos\psi\,\partial_1\theta\Big )\\
\nonumber d_{21}&=&\sqrt{g_{11}}(\hat r_3\cdot\partial_2\hat
r_1)\\ &=&
\sqrt{g_{11}}\Big(-\sin\theta\cos\psi\,\partial_2\phi
+\sin\psi\,\partial_2\theta\Big )
\end{eqnarray}
\label{eq:RoussosSecondForm}
\end{subequations}
Note that the symmetry condition $d_{12}=d_{21}$ is not explicit in
\Eq{eq:RoussosSecondForm}, but is enforced by \Eq{eq:RoussosEqs2;c}.

As an elementary illustration, we return to the case of a trumpet
characterized by a metric of the form $\sqrt{g_{11}}=\rho(v)$,
$\sqrt{g_{22}}=1$, and $g_{12}=0$. It is straightforward to verify that
  \begin{equation}
    \label{eq:RoussosTrumpet}
    \psi=0,\ \phi=\pi/2+u,\ \cos\theta=-\rho',\  \sin\theta=\sqrt{1-\rho'^2}
  \end{equation}
is a solution of \Eqs{eq:RoussosEqs2} that reproduces the trumpet
discussed earlier in this section.

Our  aim in Section \ref{sec:Evolution} is to show that \Eqs{eq:RoussosEqs2} can
be used as evolution equations actually to calculate trumpet--like
solutions for the general class of metrics given by
\Eq{eq:RoussosMetric} through the solution of an initial value
problem. The connection with the standard Gauss--Codazzi compatibility
conditions\cite{Eisenhart.59} is briefly discussed in Section \ref{sec:GaussCodazzi}.

\subsection{Evolution equations\label{sec:Evolution}}

Consider the trumpet with metric given by \Eq{eq:Trumpet}. When $v\rightarrow
-\infty$, the surface approaches a cylinder. Can one choose some large
negative value of $v$, suppose that the surface for this value of $v$
is a circle, and integrate towards $v=0$ knowing nothing but the
metric and reconstruct the trumpet? The answer is yes. One can find three
first-order equations that express the changes of $\theta$, $\psi$, and $\phi$
with respect to $v$ and integrate them forward as if $v$ is a time variable.

Finding the equations is not completely
straightforward. \Eqs{eq:RoussosEqs2} consist in three first--order
partial differential equations for the three Euler angles $\theta$,
$\psi$, and $\phi$. However since \Eq{eq:RoussosEqs2;a} involves derivatives
in $u$ only, algebraic manipulation alone does not allow one to solve
for $\partial_2\theta$, $\partial_2\psi$,  and $\partial_2\phi$. 

Note that derivatives of $\phi$ appear in \Eqs{eq:RoussosEqs2}
but not $\phi$ itself. One can use the first two of \Eqs{eq:RoussosEqs2} to
express $\partial_1\phi$ and $\partial_2\phi$ in terms of the other
two Euler angles. Removing $\phi$ in this way is only
permitted, however, if after solving for $\partial_1\phi$ and $\partial_2\phi$
one imposes the condition 
$\partial_1\partial_2\phi=\partial_2\partial_1\phi$ to obtain
%%  This
%%  compatibility condition can be imposed by differentiating
%% \Eq{eq:RoussosEqs2;a} with respect to $v$, differentiating
%% \Eq{eq:RoussosEqs2;b} with respect to $u$, subtracting the first of
%% these from the second and setting the result to zero. 
%% It looks as if
%% one has to divide through by $\cos\theta$ before carrying out this
%% procedure, but the results are the same. 
\begin{eqnarray}
\nonumber&&\big (  \sqrt{g_{11}\,g_{22}}   \partial_1\psi+\sqrt{g_{11}}
  \partial_2\sqrt{g_{11}}\big )   \partial_2\theta-\sqrt{g_{22}}G\cot\theta\\
&=&\big (\sqrt{g_{11}\,g_{22}}\partial_2\psi-\sqrt{g_{22}}
  \partial_1\sqrt{g_{22}}\big )
  \,\partial_1\theta,
\label{eq:RoussosEv0}
\end{eqnarray}
where
\begin{eqnarray}
\nonumber
G&=&{\sqrt{g_{11}}\over
  g_{22}}\,(\partial_2\sqrt{g_{11}})\,\partial_2\sqrt{g_{22}}
+{1\over \sqrt{g_{11}}} (\partial_1\sqrt{g_{11}})\partial_1\sqrt{g_{22}}\\
&-&\partial_1\partial_1\sqrt{g_{22}} -{\sqrt{g_{11}}\over
    \sqrt{g_{22}}}\partial_2\partial_2\sqrt{g_{11}}
\end{eqnarray}

One now solves the three \Eqs{eq:RoussosEqs2} and \Eq{eq:RoussosEv0} for $\partial_2\theta$, $\partial_2\psi$,
$\partial_1\phi$ and $\partial_2\phi$. In expressing the results, it
is convenient to use the expressions for the second fundamental form
in \Eq{eq:RoussosSecondForm} as shorthand for combinations of
derivatives. One obtains
\begin{subequations}
\label{eq:Roussos}
\begin{eqnarray}
\partial_2\theta&=& {d_{12}\partial_1\theta+G\cos\psi\over d_{11}} \\
\nonumber \partial_2\psi&=& {1\over d_{11}} \big[\ 
{\partial_1\sqrt{g_{22}}\over
  \sqrt{g_{11}}}\,d_{11}-\cos\theta\,d_{12}\partial_1\phi\\ 
   && ~~~~~~~~~~~~~~-G\sin\psi\cot\theta\big]\\
\partial_2\phi &=&{\partial_1\sqrt{g_{22}}\over \cos\theta
  \sqrt{g_{11}}}-{\partial_2\psi\over \cos\theta}.
 \end{eqnarray}
 These are the basic evolution equations for the Euler angles. Whenever
$\partial_1\phi$ appears, it should be viewed as shorthand for
\begin{equation}
\partial_1\phi=-{\sqrt{g_{22}}\partial_1\psi+\partial_2\sqrt{g_{11}}\over \sqrt{g_{22}}\cos\theta}.
\label{eq:Roussos;d}
\end{equation} 
\end{subequations}

The procedure for constructing surfaces progresses as follows: We
specify the values of $\theta$, $\psi$, and $\phi$ for all $u$ and
some value of $v=v_0$. To construct a trumpet, we choose $v_0$
sufficiently negative so that we can use the expressions for cylinder
$\psi=0$, $\phi=\pi/2+u$, and $\theta=\pi/2$ 
to set initial values of the Euler angles for $u\in[0,2\pi]$. Next, we calculate the right hand
sides of \Eqs{eq:Roussos} using these initial values, and update each
Euler angle through explicit Euler integration,
$\theta(v+dv)=\theta(v)+\partial_2\theta\,dv$. It is easy to construct
the surface $\vec r$. To do so, form the trihedral from the Euler
angles through \Eq{eq:eangles}. 
Then from \Eq{eq:RoussosTrihedral} one has
\begin{equation}
\partial_2\vec r=\sqrt{g_{22}}\hat r_2,
\end{equation} 
which means that if the surface is specified on the line $v=v_0$, its
future evolution can be determined as well. 

The integration process is extremely rapid. We have used it in order
to reproduce the trumpet depicted in Figure \ref{fig:Trumpet}. The
process of integrating forward terminates at $v=0$ because $d_{11}$
vanishes there for all $u$, and two denominators in \Eq{eq:Roussos} become
singular.

\begin{figure}[htbp]
\ifcolor
\epsfxsize=3.3in\epsffile{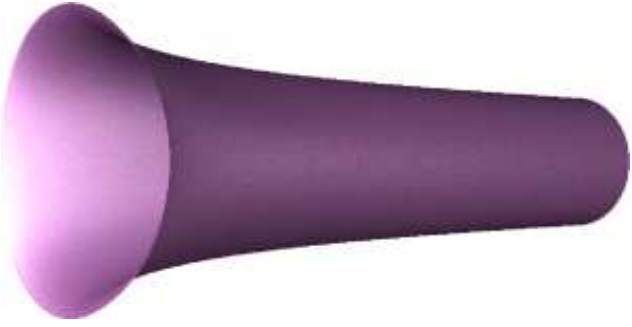}
\else
\epsfxsize=3.3in\epsffile{Roussos04-07-26__0visualize.eps}
\fi
\caption{Three--fold trumpet produced by integrating \Eqs{eq:Roussos}
  forward from a cylindrically symmetric initial condition at
  $v_0=-2\pi$, using the metric in \Eq{eq:RoussosThreeFold} \ifcolor
  (in color) \fi.}
\label{fig:Trumpet3}
\end{figure}

We have also used the equations to generate surfaces from more general
metrics, and a characteristic result is exhibited in
Figure \ref{fig:Trumpet3}. Here we chose the metric
\begin{equation}
\sqrt{g_{11}}=1+{1\over 1-v}(1+\mbox{${1\over 2}$}\, \cos 3u);\ g_{22}=1.
\label{eq:RoussosThreeFold}
\end{equation}
The surface has a slight three--fold modulation that is quite clear if
one plots Euler angles, but rather faint when one looks at the surface
itself. As in the case of the trumpet, the integration process stops near to
$v=0$. Examining the reason, one finds that once again $d_{11}$
vanishes. This three--fold trumpet does not have cylindrical symmetry,
and $d_{11}$ does not vanish simultaneously for all $u$. Instead, it
first vanishes for three values of $u$. However, the evolution
equations are singular and cannot be integrated past this point. Our
original hope had been to use evolution equations to obtain
flower--like solutions such as the one displayed in Figure
\ref{fig:Flower}. Unfortunately, the rippling edge
necessarily involves an oscillation in curvature, and $d_{11}$ must
oscillate in sign where ripples are visible. Thus our integration
procedure is intrinsically unable to obtain solutions of this
type. However, through trials with the evolution equations, we have
found that they successfully create surfaces over ranges of $u$ and
$v$ where $d_{11}$ does not vanish.

We note in closing that it is not difficult to generalize
the equations of this section to the case where $g_{12}\neq0$. One
takes
\begin{equation} 
\hat r_2={\partial_2\vec r/\sqrt{g_{22}}-(\hat
r_1/\sqrt{g_{22}}) \partial_2\vec r\cdot \hat
r_1\over \sqrt{1-g_{12}^2/g_{11}g_{22}}}
\end{equation}
and otherwise proceeds as before. The resulting expressions are
somewhat lengthy, and as we have not made use of them, we do not
record them here.

\subsection{\label{sec:GaussCodazzi}Gauss-Codazzi}

Our treatment has been based upon the compatibility conditions
\Eq{eq:RoussosCompatibility} which are formulated purely in terms of
the metric tensor $g_{\alpha\beta}$, while the second fundamental
tensor is a derived quantity displayed in \Eq{eq:RoussosSecondForm}.
By contrast, both tensors $g_{\alpha\beta}$ and $d_{\alpha\beta}$
appear as fundamental variables in the formulation of the
Gauss--Codazzi compatibility conditions employed in standard
treatments\cite{Eisenhart.59}. It would then be interesting to demonstrate
that the Gauss--Codazzi equations can actually be derived starting
from the compatibility conditions \Eq{eq:RoussosCompatibility} and the
definitions \Eq{eq:RoussosSecondForm}.

As an example, we write the first two conditions in
\Eq{eq:RoussosCompatibility} as
\begin{eqnarray}
\nonumber \partial_1\psi&=&-\cos\theta\partial_1\phi-{\partial_2\sqrt{g_{11}}\over
  \sqrt{g_{22}}} \\
\partial_2\psi&=&-\cos\theta\partial_2\phi+{\partial_1\sqrt{g_{22}}\over
  \sqrt{g_{11}}} 
\end{eqnarray}
and impose the integrability condition
$\partial_1\partial_2\psi=\partial_2\partial_1\psi$ to obtain
\begin{eqnarray}
\label{eq:Gauss}
&&\sin\theta
(\partial_1\theta\partial_2\phi-\partial_1\phi\partial_2\theta)\\
\nonumber &&~~~~= -\left
  [
\partial_1\Big({\partial_1\sqrt{g_{22}}\over\sqrt{g_{11}}}\Big)+\partial_2\Big({\partial_2\sqrt{g_{11}}
  \over \sqrt{g_{22}}}\Big) \right].
\end{eqnarray}
On the other hand, a straightforward application of definitions
(\ref{eq:RoussosSecondForm}) together with \Eq{eq:Gauss} yields
\begin{eqnarray}
d_{11}d_{22}-d_{12}d_{21}=R_{1212},\quad\quad\quad\quad\quad\quad\quad\quad\quad\quad\quad
\\
\nonumber R_{1212}=-\sqrt{g_{11}g_{22}}\left
  [
\partial_1\Big({\partial_1\sqrt{g_{22}}\over\sqrt{g_{11}}}\Big)+\partial_2\Big({\partial_2\sqrt{g_{11}}
  \over \sqrt{g_{22}}}\Big) \right],
\label{eq:Gauss2}
\end{eqnarray}
where $R_{1212}$ is indeed the correct expression for the Riemann
tensor element associated with a metric in the form of
\Eq{eq:RoussosMetric}. Thus \Eq{eq:Gauss2} reproduces the celebrated
Gauss equation.

We have not yet attempted a corresponding derivation of the two
Codazzi equations starting from our \Eqs{eq:RoussosCompatibility} and
(\ref{eq:RoussosSecondForm}) but merely state here all three
Gauss--Codazzi equations in symbolic form:
\begin{eqnarray}
\nonumber d_{22}&=&{d_{12}^2+R_{1212}\over d_{11}}\\
\partial_2 d_{11}&=& \partial_1 d_{12}+d_{1\gamma}\left\{ {\gamma\atop
  12}\right\} -d_{2\gamma}\left\{ {\gamma\atop 11}\right \} \label{eq:GaussCodazzi}\\
\nonumber \partial_2 d_{12}&=& \partial_1 d_{22}-d_{2\gamma}\left\{ {\gamma\atop
  12}\right\} +d_{1\gamma}\left\{ {\gamma\atop 22}\right \}, 
\end{eqnarray}
where $\{{\gamma\atop\alpha\beta}\}$ are the Christoffel
symbols. \Eqs{eq:GaussCodazzi} are again written in the form of
evolution equations and can be solved with initial conditions starting
at $v\rightarrow-\infty$ given the ordinary cylinder values
$d_{11}=-1$, $d_{22}=0$ and $d_{12}=0$. Once a solution of
\Eqs{eq:GaussCodazzi} is obtained at some finite $v$, the actual
construction of the surface proceeds through the solution of a
compatible system of linear equations: see Eqs. (39.8) of
Ref. \cite{Eisenhart.59}.

\section{\label{sec:Numerics}Ground state for flowers}

\subsection{Numerical technique}

We now return to techniques employed recently\cite{Marder.FOP.03} for
the construction of surfaces through minimization of an elastic energy
functional, and examine again the results in light of what we learned
in Section \ref{sec:Geometry}.  We begin by defining more carefully
than has been done previously the problem that needs to be solved. In
rough outline, one wants to take a thin flat sheet of material, impose
a new metric $g_{\alpha\beta}$ upon it, and ask how it deforms in
response. A more precise specification of the problem follows:

Differential geometry describes a mapping between two spaces: a
reference configuration described by variables $(u,v)$, and a surface
described by $\vec r(u,v)$. Experiments on deformed surfaces are
performed by taking a flat sheet of material, deforming it in some
controlled way, and then allowing the material to buckle in space.  To
construct a numerical model of the system we will first describe the
reference state, corresponding to undeformed material, and then a
discrete set of variables that describes a sheet of material moving
about in three dimensions.

The experimental reference state consists in a flat slab of material,
much wider and longer than it is thick. Imagine therefore positions in
a thin sheet described by $(x,y,z)$, where $z\in[0,t]$, and the
thickness $t$ is small. Pick $N$ points within this sheet, and label
them by $\vec r^0_i=(x_i,y_i,z_i)$, where $i$ ranges from 1 to $N$. In
practice, we will take these points to sit on regular lattices, but
they could be randomly distributed. Each point $\vec r^0_i$ has a
number of near neighbors: label these near neighbors with $j\in
n(i)$. Describe the vector between two near neighbors by $\vec
r^0_{ij}=\vec r^0_j-\vec r^0_i.$ We can now write down an
energy functional which is constructed precisely so that its ground
state gives back this reference configuration. This functional is
defined on a new collection of variables $\vec r_i$, where again $i$
ranges from $1$ to $N$, and the neighbor list $j\in n(i)$ is the same
as in the reference configuration. Now, however, the points $\vec r_i$
are free to move anywhere in three--dimensional space. One can think
of them as describing arbitrary deformations of the original thin
sheet.  If a particle $i$ has particle $j$ as a neighbor in the
reference configuration, then particle $j$ remains in the list of
neighbors no matter how the sheet deforms.

The significance of neighbors is provided by an energy functional that
depends upon the squared distance  between pairs of
neighbors. Define $\vec r_{ij}=\vec r_j-\vec r_i$; then
\begin{equation}
U_0=\mbox{${1\over 4}$}\sum_{ij}(|\vec r_{ij}|^2-|\vec r^0_{ij}|^2)^2.
\label{eq:U0}
\end{equation}
By construction this energy functional has the property that if every
particle $\vec r_i$ returns to the reference location $\vec r_i^0$,
then the energy is zero. This ground state is not unique, for the
energy is also unchanged if the locations of all the particles are
rotated and translated in three--dimensional space, as when one picks
up a piece of cardboard and translates and rotates it. Depending upon
details involving the numbers of neighbors of each particle, there may
be additional degeneracies in the ground state as well, but we will
not worry about this point right now. 

Our numerical model for deforming the sheet is to go to each bond
and stretch it so that the square distance between near neighbors $i$
and $j$ adopts the new value $l^2_{ij}$. Thus we have the energy functional
\begin{equation}
U=\mbox{${1\over 4}$}\sum_{ij}(|\vec r_{ij}|^2-l^2_{ij})^2,
\label{eq:U}
\end{equation}
and direct minimization of $U$ has been employed to obtain most of the
three--dimensional figures in this paper.

In order to connect \Eq{eq:U} with the discussion in the previous
sections, we must explain the connection between bond lengths $l_{ij}$
and metrics.  We use the following prescription. For each bond $ij$, choose
values of $u$ and $v$ through
\begin{equation}
u_{ij}=\hat e_1\cdot (\vec r_i^0+\vec r_j^0)/2; 
v_{ij}=\hat e_2\cdot (\vec r_i^0+\vec r_j^0)/2. 
\label{eq:uv}
\end{equation}
In other words, take $u$ and $v$ to be the $x$ and $y$ coordinates of
the midpoints of bonds in the reference configuration.
We define a {\it target metric} through three functions
\begin{equation}
g^t_{11}(u,v), g^t_{22}(u,v),\mbox{and}\quad g^t_{12}(u,v).
\label{eq:target}
\end{equation}
One could choose, for example the functions in \Eq{eq:TrumpetMetric}
if one wanted to recover a trumpet.  Using the target metric, one
obtains a new equilibrium length squared $l^2_{ij}$ for the bond between
points $i$ and $j$ through
\begin{equation}
l^2_{ij}=\sum_{\alpha\beta}   \vec r^{0\alpha}_{ij}
g^t_{\alpha\beta}(u_{ij},v_{ij}) \vec r^{0\beta}_{ij}.
\label{eq:lij}
\end{equation}
So, for example, if a bond in the reference configuration lies along
the $x$ axis at position $(u,v)$ given by \Eq{eq:uv}, its new length is
$\sqrt{g^t_{11}(u,v)}$. 
The reason that we call $g^t_{\alpha\beta}$ the target metric rather
than the metric is that one can put any collection of particle
locations one wishes into the functional \Eq{eq:U}, not just particle
locations that correspond to surfaces with metric
$g^t_{\alpha\beta}$. Thus the metric of the surface obtained through
numerical minimization may in principle be different from the target.
We will be attempting to determine whether the 
ground states of $U$ produce surfaces whose metric equals the target metric.

One might wonder why $U$ involves squares of bond lengths, rather than
$(r_{ij}-l_{ij})^2$. The answer is that \Eq{eq:U} leads to
conventional nonlinear elasticity in the continuum limit, while the
alternative does not. To obtain the continuum limit, let $\vec r(u,v)$
be a continuous vector field, and write
\begin{equation}
\vec r_{ij}\approx  (\vec r^0_{ij}\cdot\vec\nabla ) \vec r.
\label{eq:rij}
\end{equation}
Recalling \Eq{eq:MetricDefine},substitute \Eq{eq:rij} into \Eq{eq:U}
to obtain
\begin{equation}
U=\mbox{${1\over 4}$}\sum_{ij}\left[\sum_{\alpha\beta}   \vec r^{0\alpha}_{ij}
(g_{\alpha\beta}-g^t_{\alpha\beta}) \vec r^{0\beta}_{ij}\right]^2; 
\label{eq:U2}
\end{equation}
that is, the energy is given by subtracting the target metric from the
actual metric, and vanishes when the two are equal. From \Eq{eq:U2}
one sees that the appropriate generalization of the Lagrangean strain
tensor to situations with target metrics is
\begin{equation}
  \label{eq:LST}
  E_{\alpha\beta}=\mbox{$1\over 2$}(g_{\alpha\beta}-g^t_{\alpha\beta}).
\end{equation}
If the target metric is a unit tensor, $E$ reduces to the
conventional Lagrangean strain tensor of nonlinear elasticity, and
when deformations are small it reduces further to the strain tensor of
linear elasticity. One can write
\begin{equation}
  \label{eq:U3}
U=\sum_{ij}\left[\sum_{\alpha\beta}   \vec r^{0\alpha}_{ij}
E_{\alpha\beta} \vec r^{0\beta}_{ij}\right]^2. 
\end{equation}
For a particular arrangement of mass points, one can perform the sums
over $\vec r^0$ and obtain a specific quadratic functional depending
upon the components of $E$\cite{Marder.FOP.03}. We do not need
these expressions here are will not pursue them further.

However, we will spell out the particular reference configuration
$\vec r^0_i$ that has been used to produce results for this paper. It
consists in either one or two layers of a triangular lattice. To
be completely explicit, use three integers $lmn$ to describe
the point locations rather than the single index $i$:
\begin{equation}
\vec r^0_{lmn}=(1,0,0)l+(\mbox{$1\over 2$},\mbox{$\sqrt{3}\over
  2$},0)m+(\mbox{$1\over \sqrt{3}$},0,\mbox{$\sqrt{2\over 3}$})n,
\label{eq:reference}
\end{equation}
where $l$ and $m$ range over positive and negative integers, and $n$
equals 0 to produce a single layered structure, or 
ranges over $0$ and $1$ to produce a two--layered structure. In the
two--layered structure, each particle has nine nearest neighbors, each at
unit distance; six with the same value of $n$ on the same horizontal
sheet, and three with a different value of $n$ on a different
horizontal sheet. In these crystalline structures, the near neighbors
of particle $i$ are all the particles $j$ for which $|\vec
r^0_{ij}|^2=1$.

\begin{figure}[htbp]
  \centering
  \epsfysize=2in\epsffile{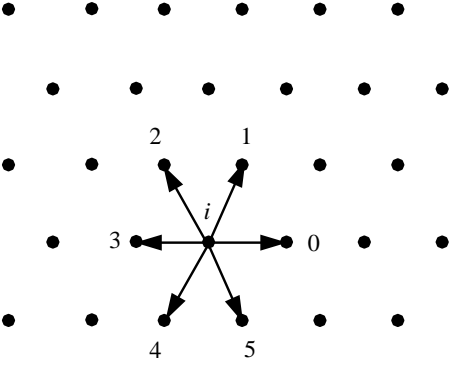}
  \caption{Enumeration of neighbors surrounding point $i$, used to
  describe the construction of numerical metrics in \Eq{eq:g_num}. The
  diagram shows the particles sitting in the reference configuration
  $\vec r_i^0$ described by \Eq{eq:reference}. } 
  \label{fig:triang3}
\end{figure}

We record one final technical point about the numerical techniques.
Given particle locations $\vec r_i$, we will want to view the
particles as describing a continuous surface, and to construct its
metric. To do so, focus on the lower sheet of particles ($n=0$ in
\Eq{eq:reference}) and label the neighbors of particle $i$ as in Figure
\ref{fig:triang3}.  Denote by $\delta r^2_0$ the square distance from
point $i$ to the neighbor located at $0$ in the figure and so on for
the remaining five neighbors. Then we have numerical representations
of the metric
\begin{eqnarray}
\nonumber g_{11}&=&{ \delta r^2_0+\delta r^2_3\over 2}\\
g_{22}&=&{\delta r^2_1+\delta r^2_2+\delta r^2_4+\delta
  r^2_5-\delta r^2_0/2- \delta r^2_3/2\over 3}\\
\nonumber g_{12}&=&{\delta r^2_1-\delta r^2_2-\delta r^2_4+\delta r^2_5\over
  2\sqrt3} 
\label{eq:g_num}
\end{eqnarray}

\subsection{Numerical experiments}

The questions we wish to pose about target metrics are:
\begin{enumerate}
\item What are the ground states of \Eq{eq:U2}?
\item When do these ground states correspond to smooth surfaces in the
  continuum limit?
\item When the ground state is a smooth surface, under what conditions
  does the target metric equal the metric of the surface?
\end{enumerate}

%% We conjecture the following
%% \begin{enumerate}
%% \item The ground states of \Eq{eq:U2} are smooth surfaces when the
%%   reference structure has some thickness and resists bending.
%% \item If a system described by \Eq{eq:U2} corresponds to one in the
%%   continuum for which the evolution equations \Eqs{eq:Roussos} can be
%%   solved, then in the ground state, the metric of the surface computed
%%   from \Eq{eq:g_num} will equal the target metric within numerical
%%   accuracy.
%% \end{enumerate}

\begin{figure}[!tbp]
  \centering
\ifcolor
  \epsfxsize=3in\epsffile{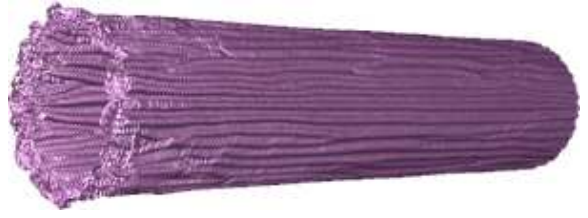}
\else
  \epsfxsize=3in\epsffile{Membrane04-07-27__1visualize.eps}
\fi
\caption{Image of surface created by minimizing \Eq{eq:U2} for a
  reference crystal with $200\times 346\times 1$ particles, and target
  metric given by \Eq{eq:g_target}. The surface achieves the target;
  the total energy summed over all particles is less than
  $10^{-2}$. However, the solution is nowhere smooth, and resembles
  tree bark. It can be understood as an embedding of the target metric
  whose first derivative is nowhere continuous \ifcolor (in color)\fi.}
  \label{fig:Bark}
\end{figure}

Observe that ground states of \Eq{eq:U2} always exist. The functional
is positive definite, and for a finite number of particles must have
one or more global minima. If the reference configuration does not
have bending stiffness, however, the resulting ground state
corresponds to a nondifferentiable surface. To
illustrate this point, we use \Eq{eq:Trumpet} as a target
metric, and work on the domain 
\begin{equation}
  \label{eq:domain}
  u\in[0,2\pi],\quad\mbox{and}\quad v\in[-3\pi,0].
\end{equation}
We represent the system with a reference crystal 200 columns long, 346
rows high, but only 1 layer thick (in \Eq{eq:reference}, $l\in[0,200],
m\in[0,346], n=0$). Since the reference configuration is infinitely
thin, there is no source of bending stiffness. Using the target metric
\begin{eqnarray}
\nonumber g^t_{11}(u,v)&=&1+{1\over 1-v}\\
g^t_{22}(u,v)&=&1\label{eq:g_target}
\\
\nonumber g^t_{12}(u,v)&=&0
\end{eqnarray}
we minimize $U$, and the result is displayed in
Figure \ref{fig:Bark}. The total energy $U$ has
converged below $10^{-2}$, and each bond has reached its target value
to better than two parts in $10^4$. However, the surface is not
smooth, nor is the configuration displayed in the figure plausibly
unique. As we know from Section \ref{sec:trumpet}, there do exist
smooth surfaces whose metric equals this target metric, but the
minimization routine under these conditions does not find
them. Venkataramani {\it et al.}\cite{Venkataramani.00} have pointed out the
existence of non--smooth surfaces of the sort appearing in the figure.

\begin{figure}[!tbp]
  \centering
\ifcolor
\epsfxsize=3in\epsffile{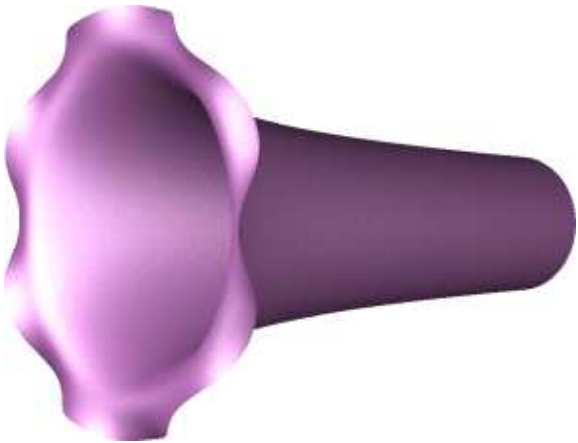}
\else
\epsfxsize=3in\epsffile{Membrane04-07-26__2visualize.eps}
\fi
\caption{Image of surface created by minimizing \Eq{eq:U2} for a
  reference crystal with $200\times 346\times 2$ particles, and target
  metric given by \Eq{eq:g_target}, corresponding to a domain where
  $v$ varies from $-3\pi$ to $1\over 2$. Once $v>0$, trumpet solutions
  given by \Eq{eq:Trumpet0} no longer exist, and the evolution
  equations in \Eq{eq:Roussos} are incapable of finding solutions. The
  minimum energy state is the smooth flower--like surface with
  seven--fold symmetry displayed here. As shown in Figure
  \ref{fig:metric_components}, the metric of this surface does not
  equal the target metric. Creation of this surface required  long
  series of minimizations. The process began by placing particles in a
  cylinder with a Euclidean metric, and very slowly changing the
  metric until it reached the desired target value, continually
  minimizing the functional \Eq{eq:U2} along the way. Attempts to
  find the surface more quickly resulted in higher--energy structures
  with creases \ifcolor (in color)\fi.}
  \label{fig:Flower}
\end{figure}

\begin{figure}
\ifcolor
\epsfxsize=\columnwidth\epsffile{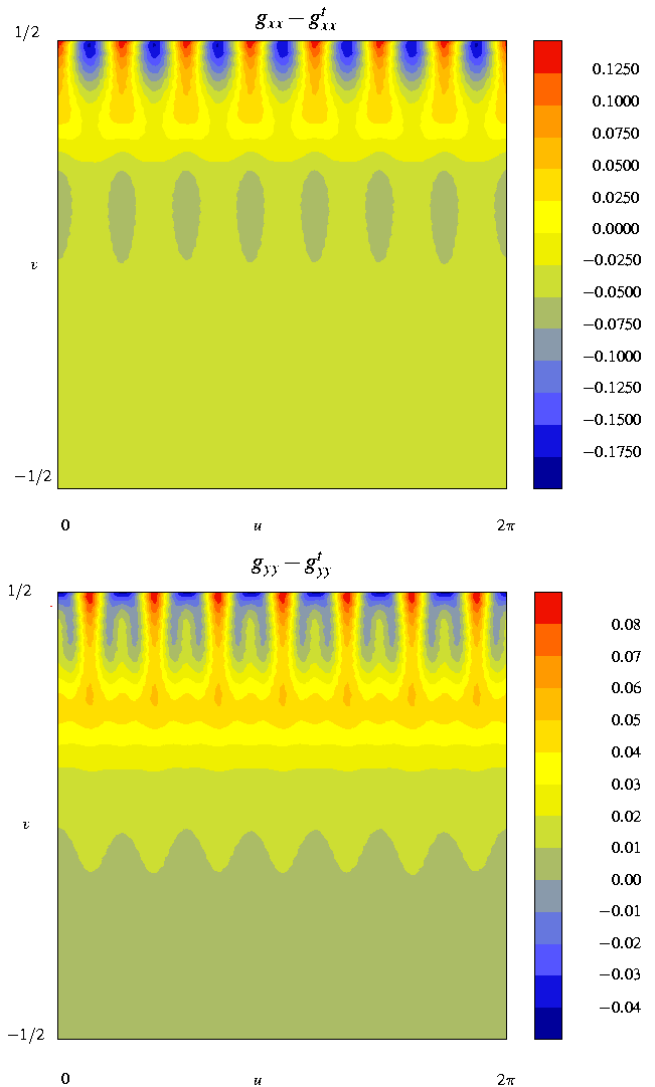}
\else
\epsfxsize=\columnwidth\epsffile{g11_g22.eps}
\fi
\caption{
Metric component deviations for the flower depicted in Figure
\ref{fig:Flower}. Each panel shows a contour plot of the difference
between the metric component $g_{\alpha\beta}$ and the target metric
$g_{\alpha\beta}^t$. The off--diagonal metric element $g_{12}$ is not
illustrated because its numerical values are too small to be discerned
at the scale of the figure \ifcolor (in color)\fi.
\label{fig:metric_components}
}
\end{figure}

Low energy states of \Eq{eq:U2} look very different when one moves
from reference crystals with one layer to reference crystals with two
layers (In \Eq{eq:reference}, $n\in[0,1]$). Now the system possesses
some stiffness, and pays an energy penalty for bending too
rapidly. Once again minimizing \Eq{eq:U2} for the parameter range in
\Eq{eq:domain}, we find the minimum energy  state that is essentially
the trumpet shown in Figure \ref{fig:Trumpet}. The energy of
this structure is 74.3; it cannot converge to zero, since springy
elastic material has been wrapped into a cylinder. 

We compared the metric $g$ as computed through \Eq{eq:g_num} with the
target metric $g^t$ for the structure in Figure
\ref{fig:Trumpet}. For each metric component, the
difference is featureless, and approximately equal to $-3\times
10^{-2}$ at every point, due to a slight compression needed to bend
the inner layer of the reference crystal into a cylinder. We conclude
from the numerical calculations that the metric and target metric
are equal within numerical accuracy. We emphasize that in Section
\ref{sec:Evolution} we reproduced the trumpet in Fig.
\ref{fig:Trumpet} through evolution equations; the metric
and boundary conditions suffice to determine the surface in this case.

As a more interesting exercise in direct minimization of \Eq{eq:U2},
we extend the 
domain beyond the critical point and look for a solution in
\begin{equation}
  \label{eq:domain2}
  u\in[0,2\pi],\quad\mbox{and}\quad v\in[-3\pi,\mbox{$1\over 2$}].
\end{equation}
The significance of increasing the range of $v$ is that the 
theory of Section \ref{sec:trumpet} is unable to find a smooth
surface for $v>0$. In contrast, upon minimizing
$U$ once more, we find the flower--like surface shown in
Figure \ref{fig:Flower}. Now when we subtract
the target metric \Eq{eq:g_target} from the metric
actually achieved, \Eq{eq:g_num}, the difference is visible, as shown
in Figure \ref{fig:metric_components}. The seven--fold pattern in the
surface is reflected in seven--fold oscillations in 
$g_{11}$ and $g_{22}$. The off--diagonal component of the tensor,
$g_{12}$, remains zero within numerical accuracy. We conclude that in
this case, no smooth surface is able to reproduce the target metric,
and numerical minimization finds a metric  close to the
target that is capable of producing a smooth surface even for
$v>0$. We see no reason 
that this surface should be considered unique. In all likelihood, its
details depend upon the thickness of the sheet. According to Audoly
and Boudaoud\cite{Audoly.03b}, one should expect the surface to be
increasingly ramified as its thickness diminishes.

As a final exercise, we took the metric components computed
numerically and depicted in Fig. \ref{fig:metric_components}, inserted
them into the evolution equations \Eqs{eq:Roussos} and attempted to
reproduce the structure in Figure \ref{fig:Flower}. This
attempt was only partially successful. The evolution equations were unable to
proceed past points where $d_{11}$ approached zero. It is clearly
possible for $d_{11}$ to vanish without a surface developing singular
cusps at that point, but our numerical routines would require very
delicate cancellations in order to proceed past such a point.

\section{Conclusions}

Given a surface, it is completely straightforward to compute the
metric. The inverse problem of finding surfaces compatible with a
given metric is much more difficult. In setting out on the studies
recorded in this paper, we had two goals. First, we wanted finally
to determine the low--energy configurations of long strips with linear
gradients in metric. This we have accomplished. Second, we hoped to
determine conditions on the relatively simple metrics thought to
create the shapes of leaves and flowers\cite{Sharon.02} that would
enable surfaces to be reconstructed.

In the second task we have been only partially successful. We found
evolution equations enabling construction of surfaces from initial
conditions and metric alone. However, at points when $d_{11}$, a
component of the second fundamental form, vanishes, the equations
become singular and cannot be integrated further. We are unable to
tell when this singularity really reflects the impossibility of
creating a surface compatible with the metric, and when it simply
reflects a technical defect in the method of construction.

Some intuitive understanding of this situation can be obtained by
thinking about a piece of paper, with flat Euclidean metric. Imagine
holding the bottom of the paper completely straight, along the
$u$ axis, while the left edge of the paper runs along the $v$
axis. There is an infinite number of shapes the paper can take, which
correspond to all different bends possible at different points along
$v$ around axes parallel to $u$. All of these solutions have vanishing
curvature in the $u$ direction; $d_{11}=0$, and the evolution
equations (\ref{eq:Roussos}) accordingly are unable to make any
predictions. That is, in some cases the vanishing of $d_{11}$ can
correspond to a genuine uncertainty, based upon initial conditions,
concerning how the surface should evolve.

On the other hand, for the surface in Figure \ref{fig:Flower}, the
curvature $d_{11}$ oscillates in sign and therefore passes through
zero, yet the surface exists.  Therefore, the vanishing of $d_{11}$
can be compatible with the existence of a surface even if our purely
geometrical methods cannot find it beyond points where $d_{11}=0$.
Minimization of \Eq{eq:U} still produces a surface because elasticity
resolves the questions that cannot be answered by geometry.

\acknowledgements Most of this work was carried out within the
stimulating environment of the Roussos Center for Nonlinear Dynamics,
and was influenced in part by the Hydrangeas at Boukari,
Corfu. M.M. is grateful for financial support from the National
Science Foundation, DMR-0401766, and the Research Center of Crete.

\bibliography{/home/marder/crack/tex/membrane/membrane,/home/marder/grants/marder}
\end{document}